# Space Time Reflection Symmetry in the Jones Formalism in Optics

Jean-François Bisson

Département de physique et d'astronomie, Université de Moncton

[Jean-francois.bisson@umoncton.ca](mailto:Jean-francois.bisson@umoncton.ca)

We provide a description of the parity space reflection-time reversal (*PT*) operator acting on the two-dimensional polarization space of light represented by the linear algebra of Jones vectors and matrices. We establish the form of a *PT*-symmetric Jones matrix. We present two examples of laser resonators whose polarization eigenstates are described by *PT*-symmetric Jones matrices: one is based on the Faraday effect and a dichroic attenuation, while the other is made of twisted anisotropic mirrors. Both possess a control parameter that experimentally covers the exact, the broken *PT*-symmetry regions and their boundary, called an exceptional point, where the eigenstates of the resonator coalesce into a single state. The exact *PT*-symmetric region produces laser polarization modes emitting at the same frequency with different intracavity losses, while the broken *PT*-symmetric region features polarization modes emitting at distinct frequencies with the same intracavity losses. By applying unitary transformations, the concept of *PT*-symmetric Jones matrix is extended to matrices that commute with any antiunitary operator, thereby opening the prospect of a larger family of resonators geometries that also feature real or complex-conjugate spectra.

# 1. INTRODUCTION

Symmetries are operations that leave a physical system unchanged; they reveal conservation laws, they enable a labeling and classification of eigenstates of a quantum system, they determine their degeneracy and the selection rules for transitions between pairs of energy levels of a quantum system interacting with an external perturbation [1-3]. One of the most fundamental symmetries of nature is probably the combined charge conjugation, parity and time-reversal (*CPT*) symmetry, which states that every quantum system has a symmetry operation that simultaneously takes the anti-particle, reverses the space coordinates and the direction of time [4,5]. Lorentz invariance and a positive energy spectrum are required for establishing *CPT* invariance in quantum field theory [5,6]. The hermiticity of the Hamiltonian, *H*, which is an operator equal to its adjoint, i.e., $H^\dagger = H$, is generally presented as a prerequisite for obtaining such a real energy spectrum bounded below [7].

Bender and Boettcher, questioning this requirement, postulated instead that a Hamiltonian could possess space-time (*PT*) reflection symmetry [8]. They showed that such operator can exhibit, like Hermitian systems, an entirely real eigenvalue spectrum. In addition, the authors constructed an operator *C* [9], similar to the charge conjugation operator, that allowed them to define a positive norm of the quantum state that is constant in time, which is required for the conservation of probability of a closed system. Doing so, the authors replaced the hermiticity requirement of *H* by another one, apparently more grounded into the symmetry of Nature. Now, when varying the value of a control parameter, *PT*-symmetric Hamiltonians, $H_{PT}$, feature a transition between a so-called *unbroken* or *exact* symmetry region, where the eigenvalue spectrum is entirely real and the eigenstates of $H_{PT}$ are shared with operator *PT*, and a *broken* symmetry region, where the spectrum is no longer real and eigenstates of $H_{PT}$ differ from those of *PT*. At the transition between the exact and broken symmetry regions, some eigenstates and eigenvalues coalesce into a single state, and upon entering the broken symmetry region, the eigenvalue splits into a pair of complex conjugates [8,9].

Although the concept of a *PT*-symmetric Hamiltonian has found limited applications in quantum mechanics so far, it has been widely applied in the field of optics [10-12]. In waveguide optics, it is well known that the equation describing the propagation of coherent light in the paraxial approximation is isomorphic to the Schrödinger equation, with time *t* replaced by the propagation direction *z*, the wavefunction by the envelope of the electric field, and the potential *V* by the refractive index, such as $V(x) \leftrightarrow \frac{-2\pi\hbar}{\lambda} n(x)$, where *x* is a transverse coordinate. In a PT-symmetric Hamiltonian, the complex potential has the symmetry: $V(x) = V^*(-x)$ [13]. In coupled optical waveguides, this can be mimicked by making the real part of the refractive index equal in both waveguides, and opposite values of its imaginary part, i.e., gain in one waveguide perfectly balanced by loss in the other. The transition between exact and broken symmetry in such photonic structures was used to produce original effects, such as loss-induced transparency in passive waveguides, non-reciprocal light transmission, unidirectional reflectance, coherent perfect absorption and for achieving single mode laser oscillation [10-12].

It was also shown that *PT*-symmetric lasers do not require the interaction between coupled waveguides or resonators and can be realized by harnessing *PT* symmetry in the two-dimensional space of polarization [14,15]. In substance, a resonator made of a pair of anisotropic mirrors with suitable optical characteristics can display *PT* symmetry; the relative orientation of the two mirrors' principal axes can be used as a control parameter that spans regions of exact and broken *PT* symmetry, including the transition at the branch point singularity where the two states become degenerate. Now, unlike the *PT* symmetry applied to the Schrödinger equation, which uses a Hilbert vector space of high dimension, the polarization vector space only has two dimensions. This makes it easier to theoretically describe *PT* symmetry and to realize it in the laboratory. To our knowledge, the topic of *PT* symmetry in the polarization space has only been skimmed over in the literature, with the demonstration of exceptional points in the polarization space in metasurfaces or thin films with [16] or without [17-18] reference to *PT* symmetry.

In this paper, we define the *P* and *T* operators of the polarization state and provide their representations in different orthogonal bases of polarization eigenstates. We introduce the concept of a *PT*-symmetric Jones matrix, $J_{PT}$, i.e., a Jones matrix that commutes with the *PT* operator, and then provide its form in different bases. We emphasize the antilinearity of the *PT* operator, which is responsible for the possibility that, although $J_{PT}$ commutes with *PT*, they do not share common eigenstates: this is the so-called *broken* PT symmetry. We establish a criterion for a *PT*-symmetric matrix to share common eigenvectors with PT, which corresponds to *exact PT* symmetry, in which case the eigenvalues of $J_{PT}$ are real. We show that, at the transition between the two regions, both the eigenvalues and eigenvectors coalesce into a single entity. Next, we present two architectures of *PT*-symmetric laser resonators that possess a control parameter that allows one to scan the exact and broken *PT*-symmetry regions but differ by the kind of optical anisotropy involved. We show that these two designs differ in that only one resonator enables the elimination of spatial hole burning, which is key to obtaining single mode emission in microchip lasers. Finally, by using unitary transformations of Jones matrices, the concept of *PT* symmetry is shown to be a special case of a more general family of matrices, dubbed *P'T'*-symmetric, that commute with any antiunitary operator. Such *P'T'*-symmetric Jones matrix exhibits the same features of purely real and complex conjugate pairs of eigenvalues separated by an exceptional point. We conclude by showing that Hermitian and unimodular unitary matrices, which respectively describe polarizers and retarders, form special cases of *P'T'*-symmetric matrices.

## 2. THE PARITY-REFLECTION TIME REVERSAL OPERATOR IN POLARIZATION SPACE

The spatio-temporal dependence of the electric field vector, $\vec{E}$, of an ideal coherent monochromatic electromagnetic plane wave propagating in the *z* direction can be written as [19]:

$$\vec{E} = E_{0x}\cos(kz - \omega t + \varphi_x)\hat{i} + E_{0y}\cos(kz - \omega t + \varphi_y)\hat{j}, \tag{1}$$

where $E_{0x}$ and $E_{0y}$ are the amplitude of the $x$ and $y$ components of the electric field, $\hat{i}$ and $\hat{j}$ are unit vectors in the $x$ and $y$ directions, $\varphi_x$ and $\varphi_y$ are phase values of the two components of the electric field, $k$ is the modulus of the wavevector, $\omega$ is the angular frequency and $t$ is time. At any given $z$ value, the tip of $\vec{E}$ usually draws an ellipse as a function of time, as shown in Fig. 1. The values of $E_{0x}$, $E_{0y}$ and phase difference $\Delta\varphi \equiv \varphi_y - \varphi_x$ determine the orientation of the ellipse, its eccentricity, and the direction of rotation of the electric field with respect to the direction of the wavevector $\vec{k}$ [19]. The state of polarization can be described in condensed notation by using two-dimensional vectors of complex numbers called Jones vectors [20]. For instance, the field in eq. (1) can be described, in the basis of horizontal and vertical polarizations, as:

$$\vec{u}_{\{h,v\}} = \left(E_{0x}\exp(i\varphi_x), \ E_{0y}\exp(i\varphi_y)\right)^T, \tag{2}$$

and the actual field, $\vec{E}$, can be retrieved by:

$$\vec{E} = \text{Re}\left[\vec{u}\exp(i(kz - \omega t))\right]. \tag{3}$$

The effect of passive optical polarizing elements on Jones vectors can be obtained by using two-dimensional square matrices called Jones matrices [20].

Let us define the parity-reflection and time-reversal operators, $P$ and $T$, acting on the polarization state as follows: The $P$ operator consists in switching the sign of all three spatial coordinates. Within a $\pi$ rotation, it is equivalent to reflecting the light with a standard mirror or taking the mirror image of the polarization ellipse through a vertical plane, Fig. 1a. The time reversal operator $T$, which is in fact better called the *motion-reversal* operator, flips both the direction of rotation and the orientation of $\vec{k}$, while keeping the same elliptical shape, Fig. 1b. Finally, the combination of $P$ and $T$ is equivalent to reversing the direction of rotation of the electric field without changing the ellipse of polarization nor the direction of $\vec{k}$, Fig. 1c.

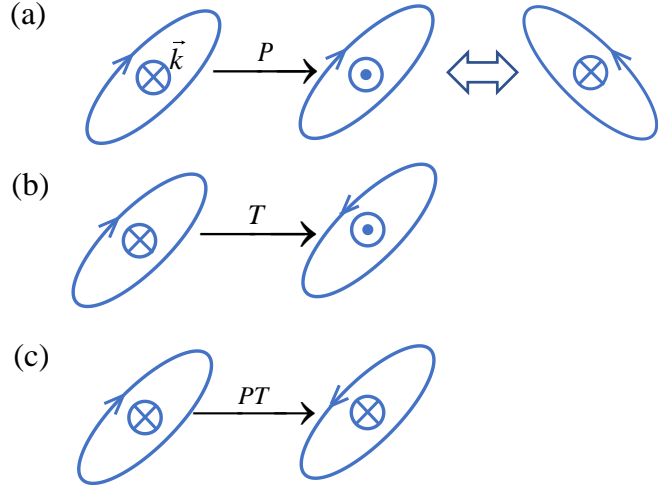

Fig. 1 Definitions of the a) parity (*P*) and b) time reversal (*T*) operators. c) The effect of applying consecutively both *P* and *T* on a given polarization state.

Can *P*, *T*, and *PT* be represented by Jones matrices? For *P*, the answer is yes. The exact form it takes depends in which basis of (generally orthonormal) vectors it is expressed. In the basis of horizontal, vertical polarization states $\{h,v\}$, oriented in the positive *x*, *y* directions, it is:

$$P_{\{h,v\}} = \begin{pmatrix} -1 & 0 \\ 0 & 1 \end{pmatrix}. \tag{4}$$

A normal isotropic mirror produces exactly this, provided the Jones vectors of the incoming and the reflected beam are expressed with right-handed *x*, *y*, *z* coordinate axes with the *z* axis always pointing in the direction of the ***k*** vector, the vertical *y* axis is the same and the horizontal *x* axis is reversed to obtain right-handed coordinates axes. This operator can be expressed in any orthogonal basis $\{\vec{u}_1, \vec{u}_2\}$ by using a unitary transformation, such as:

$$P_{\{u_1,u_2\}} = R^{-1}_{\{h,v\} \to \{\vec{u}_1,\vec{u}_2\}} P_{\{h,v\}} R_{\{\vec{u}_1,\vec{u}_2\} \to \{h,v\}}, \tag{5a}$$

where *R* is a change-of-basis unitary matrix. In this work, we focus our attention on the left and right circular polarization states, $\{l, r\}$, in which *P* is expressed as:

$$P_{\{l,r\}} = R^{-1} P_{\{h,v\}} R = \begin{pmatrix} 0 & 1 \\ 1 & 0 \end{pmatrix}, \tag{5b}$$

where $R = \frac{1}{\sqrt{2}} \begin{pmatrix} 1 & -1 \\ i & i \end{pmatrix}_{\{l,r\} \to \{h,v\}}$ is the transformation matrix from $\{l,r\}$ to $\{h,v\}$ basis.

As for the motion reversal operator *T*, it consists in reversing the arrow of time, so the direction of rotation of the electric field is reversed, which means that it also reverses the angular momentum of any state of polarization. This cannot be obtained with passive linear elements, although nonlinear optical interactions can be used to achieve this function [21-23]. Hence, $T$ cannot be described merely by Jones matrices. However, it can be described by a combination of a Jones matrix and the complex conjugate operator, $C$. In the $\{h,v\}$ basis, it can be expressed as:

$$T_{\{h,v\}} = \begin{pmatrix} 1 & 0 \\ 0 & -1 \end{pmatrix} C, \tag{6}$$

while in the $\{l,r\}$ basis, it is simply:

$$T_{\{l,r\}} = C, \tag{7a}$$

In general, for any orthogonal basis of polarization states, $\{\vec{u}_1, \vec{u}_2\}$, $T$ is expressed as:

$$T_{\{u_1,u_2\}} = R^{-1}_{\{l,r\}\to\{u_1,u_2\}} T_{\{l,r\}} R_{\{u_1,u_2\}\to\{l,r\}} = R^{-1}CR = R^{-1}R^*C, \tag{7b}$$

where $*$ stands for complex conjugate. Because $R$ is unitary, so is $R^{-1}R^*$. Hence, in a given basis of orthogonal Jones vectors, the $T$ operator is expressed by a unitary matrix, $U$, multiplied by the complex conjugate, i.e.,

$$T_{\{u_1,u_2\}} = UC. \tag{8}$$

The complex conjugation operator appearing in $T$ makes it antilinear, i.e., an operator $\Theta$ is antilinear if [1-3]:

$$\Theta(c_1\vec{u}_1 + c_2\vec{u}_2) = c_1^*\Theta(\vec{u}_1) + c_2^*\Theta(\vec{u}_2). \tag{9}$$

Eq. (8) has the same form as the time reversal operator in quantum mechanics [1-4]. The unitarity condition on $U$ guarantees the conservation of the probability of finding the particle under time reversal [1-3]. In quantum mechanics, just like in optics, the $U$ matrix depends on the considered representation. For example, the time-reversed wave function is given by:

$$\psi(x) \xrightarrow{T} \psi^*(x), \tag{10a}$$

in the $x$ representation, while it is slightly more complicated in the $p_x$ representation [2];

$$\phi(p_x) \xrightarrow{T} \phi^*(-p_x). \tag{10b}$$

In addition, in quantum systems, the square of $T$ is either the identity, $I$, or minus the identity, $-I$, depending on whether the angular momentum quantum is integral or half integral, respectively [1-3]. In the polarization space, we always have $T^2=I$, which is consistent with the fact that photons are bosons with integral angular momentum.

The $PT$ operator is formed by combining operators $P$ and $T$ and this also makes $PT$ an antilinear operator. In addition, by inspection of Fig. 1 or from eqs. (5b) and (7a), we see that $P$ and $T$ commute:

$$PT = TP, \tag{11a}$$

and that PT is involutive, i.e.,

$$(PT)^2 = I. \tag{11b}$$

### 3. PT-SYMMETRIC JONES MATRICES, $J_{PT}$

3.1 The definition and the form of $J_{PT}$

A Jones matrix, $J_{PT}$, is PT-symmetric if and only if it commutes with PT:

$$[J_{PT}, PT] \equiv J_{PT} PT - PT J_{PT} = 0. \tag{12}$$

Let us calculate $J_{PT}$ in the $\{h,v\}$ basis; from eq. (4) and (6), the PT operator then takes the simple form:

$$PT_{\{h,v\}} = -C. \tag{13}$$

The minus sign in the right-hand side is a global phase factor that has no important physical meaning. Let us assume that:

$$J_{PT\{h,v\}} = \begin{pmatrix} a & b \\ c & d \end{pmatrix}, \tag{14}$$

where all matrix elements are a priori arbitrary complex numbers. Then, from (13-14) we have:

$$PT J_{PT} = \begin{pmatrix} -a^* & -b^* \\ -c^* & -d^* \end{pmatrix} C = J_{PT} PT = \begin{pmatrix} -a & -b \\ -c & -d \end{pmatrix} C, \tag{15}$$

and we conclude that $a$, $b$, $c$, and $d$ are real numbers with no further restriction. In the $\{l,r\}$ basis, $J_{PT}$ takes the form:

$$J_{PT\{l,r\}} = \begin{pmatrix} \beta + i\gamma & \delta + i\eta \\ \delta - i\eta & \beta - i\gamma \end{pmatrix}, \tag{16}$$

where $\beta$, $\delta$, $\gamma$ and $\eta$ are real. In any given representation, $J_{PT}$ has four free parameters. The most general expression of $J_{PT}$ in any orthogonal basis of Jones vectors has six free parameters and can be expressed as [24,25]:

$$J_{PT} = \begin{pmatrix} A + iB\cos\theta + C\sin\theta & (-iB\sin\theta + C\cos\theta + iD)\exp(i\varepsilon) \\ (-iB\sin\theta + C\cos\theta - iD)\exp(-i\varepsilon) & A - iB\cos\theta - C\sin\theta \end{pmatrix}, \tag{17}$$

where $A$, $B$, $C$ and $D$ are real, $\theta \in [0, \pi]$, and $\varepsilon \in [0, 2\pi]$. We will see that Hermitian matrices are a particular case of *PT*-symmetric matrices.

3.2 Spontaneous breaking of PT symmetry and the existence of exceptional points.

Commuting operators play an important role in quantum mechanics because they correspond to compatible physical quantities that can be measured in any order without changing the physical predictions. Operators commuting with the Hamiltonian also specify conserved quantities of a quantum eigenstate. This property originates from a general theorem of linear algebra stating that any two commuting diagonalizable operators, *M* and *N*, can be simultaneously diagonalized and thus share a common basis of eigenvectors [26]. If *M* or *N* is not diagonalizable but still commute, then *M* and *N* share at least one eigenvector. Hermitian operators have orthogonal eigenvectors; their associated matrix is thus diagonalizable and commuting operators can share the same eigenvectors.

Now, following an argument by Bender et al. [9], let us suppose that the two commuting operators $J_{PT}$ and *PT* share an eigenvector, $\vec{u}$, i.e.,

$$PT\vec{u} = \lambda \vec{u} \tag{18}$$

and

$$J_{PT}\vec{u} = \Lambda \vec{u}. \tag{19}$$

Since *PT* acting on a scalar consists in taking the complex conjugate, it is thus possible to define the global phase factor of $\vec{u}$ to force the eigenvalue $\lambda$ of *PT* to be real. From eq. (13), we immediately see that the eigenvectors of *PT* are:

$$\vec{u}_{\{h,v\}} = \begin{pmatrix} c_1 & c_2 \end{pmatrix}^T, \tag{20}$$

where $c_1$ and $c_2$ are any real numbers. Hence, the eigenstates of *PT* are any rectilinear polarization state, and $\lambda = 1$. Now we have:

$$PTJ_{PT}\vec{u} = PT\Lambda\vec{u} = \Lambda^* \lambda \vec{u} \tag{21}$$

and:

$$J_{PT}PT\vec{u} = J_{PT}\lambda\vec{u} = \Lambda\lambda\vec{u}. \tag{22}$$

Because the two operators commute, (21) and (22) are equal and we conclude that $\Lambda$ is real. From the theorem outlined above, its seems that we have expanded the condition of the reality of the eigenvalues from Hermitian to a larger set of matrices: those that are *PT*-symmetric.

But is it true? Let us look at $J_{PT}$ in the particular $\{h,v\}$ basis, eq.(14). In that basis, the eigenvalues of $J_{PT}$ are:

$$\Lambda_\pm = \frac{a+d}{2} \pm \sqrt{\left(\frac{a-d}{2}\right)^2 + bc} \tag{23}$$

and the eigenvectors are:

$$\vec{u}_{\pm\{h,v\}} = \begin{pmatrix} b \\ \Lambda_\pm - a \end{pmatrix}. \tag{24}$$

We can see that the eigenstates of $J_{PT}$ are shared by $PT$ (cf. eq. (20)) only if the discriminant is positive, $\Delta = (a-d)^2/4 + bc \geq 0$, in which case the eigenvalues are indeed real. However, for $\Delta<0$, the polarization eigenstates of $J_{PT}$ are in general elliptical, and thus not shared with those of $PT$, and then the eigenvalues are a pair of complex conjugate numbers.

When the eigenvectors of $PT$ and $J_{PT}$ are shared, i.e., for $\Delta\geq 0$, the eigenvalue spectrum is purely real and the $PT$ symmetry is said to be *exact* or *unbroken*. Otherwise, the PT symmetry is said to be *broken*. At the transition between the two regions, $\Delta=0$, $J_{PT}$ becomes defective, i.e., it is no longer diagonalizable. This point is called an exceptional point (EP) [27]. At the EP, the pair of polarization eigenstates merge into a single entity and so do the eigenvalues. Therefore, the theorem stated above, that commuting linear operators share common eigenvectors, does not apply to $J_{PT}$ and $PT$. This peculiar behavior takes its origin in the fact that the theorem only applies to linear operators, whereas here, one operator, $PT$, is antilinear.

These results were derived for a particular basis of eigenvectors, but they are valid in general, independently of the chosen basis. This is the consequence of an important theorem that states that unitary equivalent observables have identical spectra [2].

It will be convenient in the next section to express the conditions of exact symmetry in the $\{l,r\}$ basis, which is:

$$\chi \equiv \gamma^2 / (\delta^2 + \eta^2) \leq 1, \tag{25}$$

broken symmetry arising if:

$$\chi > 1. \tag{26}$$

The reason for using $\chi$, which is defined in the context of the $\{l,r\}$ basis, instead of the more natural discriminant, $\Delta$, is that Hermitian matrices are special cases of PT-symmetric matrices and they all correspond to a specific value of $\chi=0$, whereas they can take different positive $\Delta$ values. Note that, in contrast to Hermitian matrices, the eigenvectors of PT-symmetric matrices are in general non orthogonal.

# 4. PT-SYMMETRIC LASER RESONATORS IN THE POLARIZATION SPACE

Polarization eigenstates play an important role in laser science. In many applications, laser sources emitting at a single frequency are needed. These includes optical sources for telecommunications [28], non-linear optics [29-31], metrology [32-33] and more recently lidars for autonomous vehicles [34]. Even when the laser operates in a single transverse, longitudinal mode, dual polarization may produce dual frequency emission because of residual anisotropy that lifts the degeneracy between the two polarization eigenstates. Hence, pure polarized emission is important for sources emitting at a single frequency. *PT*-symmetric lasers in the polarization space will be defined as laser resonators whose round-trip Jones matrices are *PT*-symmetric. We will see that they offer the possibility to eliminate dual polarization at the root by operating at an EP, where only one polarization eigenstate exists. We will also see that one kind of *PT*-symmetric lasers is also useful for suppressing multiple longitudinal mode emission, by cancelling the intensity contrast of the standing wave, thereby eliminating axial spatial hole burning (SHB). Otherwise, when SHB takes place, axial modes other than the dominant one can tap into unsaturated, high-gain regions of the active medium where destructive interference of the counter-propagating waves takes place, thereby allowing these other axial modes to oscillate [35,36].

Here, we present two designs of *PT*-symmetric resonators. They both require a combination of diretardance (i.e., direction-dependent phase shift) and diattenuation (i.e., direction-dependent dichroism) and both also possess a control parameter that allows one to scan the broken and exact symmetry regions and probe the vicinity of an EP as well. These laser architectures [37, 38] were previously introduced to study the excess quantum noise, also known as Petermann noise factor [39, 40], produced when two modes of a resonator, either with different transverse or longitudinal indices or polarization states, merge into a single state. Their *PT*-symmetric character however had remained unnoticed until recently. In the following, we analyze each resonator from its round-trip Jones matrix, $J_{RT}$. Before that, we provide an interpretation for the eigenvectors and eigenvalues of this matrix.

### 4.1 The interpretation of the eigenvectors and eigenvalues of $J_{RT}$

In the calculation of $J_{RT}$, we neglect the gain inside the active material and assume that the latter is isotropic. Let $\vec{u}^{[n]}$ be the state of polarization of the wave propagating in some direction at a fixed position inside the resonator after the $n^{\text{th}}$ round trip. After the next round-trip, it becomes:

$$\vec{u}^{[n+1]} = J_{RT}\vec{u}^{[n]}. \tag{27}$$

According to (3), the electric field is transformed as:

$$\vec{E}^{[n+1]} = J_{RT}\vec{E}^{[n]}\exp(-i\omega T_{RT}), \tag{28}$$

where $T_{RT}$ is the round-trip time of the resonator. Let $\vec{U}_+$ and $\vec{U}_-$ be the eigenvectors of $J_{RT}$, with respective eigenvalues $\Lambda_\pm \triangleq |\Lambda_\pm|\exp(\pm i\varphi)$, with $\varphi=0$ in the exact and $|\Lambda_+|=|\Lambda_-|$ in the broken symmetry regions. We then have:

$$\vec{U}_\pm^{[n+1]} = \Lambda_\pm \exp(-i\omega T_{RT}) \times \vec{U}_\pm^{[n]}. \tag{29}$$

We note that the magnitude of the eigenvalues indicates the modal round-trip losses of the passive cavity due to the cavity mirrors. Their phase is connected to the oscillation frequency: when oscillation takes place, the round-trip phase shift must be an integral multiple of $2\pi$, $p$; for the same modal transverse and longitudinal indices, we have for the two polarization eigenstates:

$$\varphi - \omega_+ T_{RT} = -\varphi - \omega_- T_{RT} = 2\pi p. \tag{30}$$

Hence, in the exact symmetry region, the oscillation frequencies are the same because $\varphi=0$, while in the broken symmetry region, the oscillation frequencies of the two modes differ by:

$$\nu_+ - \nu_- = \Delta\nu_{FSR} \times \varphi/\pi, \tag{31}$$

where $\Delta\nu_{FSR} = T_{RT}^{-1}$ is the free spectral range of the resonator.

It is worth noting that the structure of the eigenvalue spectrum of *PT*-symmetric Jones matrices differs from that of a two-level quantum system or its equivalent in optics. The exact *PT*-symmetric phase in the latter generally corresponds to distinct emission frequencies but equal dissipation in the two eigenmodes: this is exactly the opposite to *PT*-symmetric Jones matrices of a laser resonator in the exact *PT*-symmetric phase, where the two modes have the same emission frequency but different losses. The reversed situation takes place in the broken *PT*-symmetric phase. The reason for this is the presence of the imaginary factor *i* in Schrödinger's equation, which is absent in the equivalent differential equation in the Jones formalism, as outlined in Appendix A.

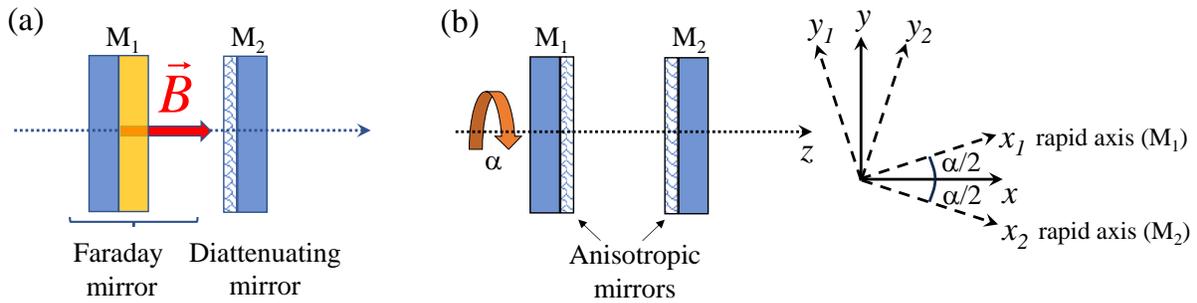

Fig. 2. Two designs of *PT*-symmetric lasers. a. The dichroic Faraday resonator has a non-reciprocal mirror, $M_1$, that displays the Faraday effect and a dichroic mirror, $M_2$, with two orthogonal principal axes. An external magnetic field is used as a control parameter to scan the exact and broken symmetry region. b. The twisted-mode resonator has two mirrors that feature both a $\pi$ phase shift and diattenuation in orthogonal axes. The rotation angle $\alpha$ between the two mirrors' principal axes, shown at the right of Fig. 2b, is the control parameter.

### 4.2 The dichroic Faraday resonator

The first *PT*-symmetric design, shown in Fig. 2a, involves a resonator with a non-reciprocal mirror, $M_1$, that presents the Faraday effect and a diattenuating mirror, $M_2$, with different reflectance values in orthogonal axes but no diretardance. Upon reflection at $M_1$, the right and left-circular polarization states are swapped, like an ordinary mirror, but a phase shift, $2\varphi$, proportional to the magnetic field component parallel to the $z$ axis, $B_z$, is produced between different circularly-polarized states. In the $\{l,r\}$ basis, the Jones matrix of the Faraday mirror, $M_1$, is then given by:

$$M_1 = \begin{pmatrix} 0 & \exp(i\varphi) \\ \exp(-i\varphi) & 0 \end{pmatrix}_{\{l,r\}}. \tag{32}$$

The Jones Matrix of the dichroic mirror, $M_2$, is given by:

$$M_2 = \begin{pmatrix} r_1 & 0 \\ 0 & -r_2 \end{pmatrix}_{\{h,v\}} = \begin{pmatrix} \Delta r & r \\ r & \Delta r \end{pmatrix}_{\{l,r\}}, \tag{33}$$

where $r_1$ and $r_2$ are (different) real positive reflection coefficients, $\Delta r \equiv \frac{1}{2}(r_1 - r_2)$ and $r \equiv \frac{1}{2}(r_1 + r_2)$. Assuming that the active medium is isotropic, the round-trip Jones matrix of this resonator, $J_F$, is given by:

$$J_F = M_1 M_2 = \begin{pmatrix} r\exp(i\varphi) & \Delta r\exp(i\varphi) \\ \Delta r\exp(-i\varphi) & r\exp(-i\varphi) \end{pmatrix}_{\{l,r\}}. \tag{34}$$

$J_F$ has the form of eq. (16), so it is *PT*-symmetric, and the control parameter, $\chi$, defined in (25), is given by:

$$\chi = r^2 \sin^2 \varphi(B_z) / \Delta r^2. \tag{35}$$

Hence, the amplitude of the $z$-component of the magnetic field serves as a control parameter that allows us to scan the exact and broken symmetry regions. At zero field, $\varphi=0$, $J_F$ is Hermitian, and the polarization eigenstates are orthogonal and rectilinear; at low field values ($\chi<1$), the eigenstates remain rectilinear but rotates towards each other as the magnetic field increases; they eventually merge into a unique rectilinear state bisecting the $x$ and $y$ axes at the EP for $r\sin\varphi(B_z) = \pm\Delta r$, then they become elliptical at higher field values and the eccentricity of the elliptical polarization increases until $\sin(\varphi)=\pm 1$ and then goes down again at even higher field, etc. Numerical simulations showing this behavior are shown in Fig. 3c.

### 4.3 The twisted mode resonator with diattenuation

The second example of a *PT*-symmetric resonator is the diattenuating twisted-mode resonator, in which both mirrors have a π phase shift diretardance between two orthogonal principal axes and at least one mirror shows diattenuation with respect to the same principal axes, i.e.,

$$M_1 = \begin{pmatrix} r_{11} & 0 \\ 0 & r_{12} \end{pmatrix}_{\{h,v\}} = \begin{pmatrix} r_1 & \Delta r_1 \\ \Delta r_1 & r_1 \end{pmatrix}_{\{l,r\}} \tag{36}$$

and

$$M_2 = \begin{pmatrix} r_{21} & 0 \\ 0 & r_{22} \end{pmatrix}_{\{h,v\}} = \begin{pmatrix} r_2 & \Delta r_2 \\ \Delta r_2 & r_2 \end{pmatrix}_{\{l,r\}}, \tag{37}$$

with similar definition for $r_i$ and $\Delta r_i$ as for the Faraday resonator. Note that positions of $r_i$ and $\Delta r_i$ in the matrix are exchanged compared to the previous case because of the π phase shift in the present case. One mirror is twisted by an *α* angle with respect to the other. Assuming an isotropic gain medium, the round-trip Jones matrix, $J_{TM}$, defined in the $\{h,v\}$ basis defined by the $x,y$ axes shown in Fig. 2b, is:

$$J_{TM} = \Theta M_1 \Theta \Theta M_2 \Theta, \tag{38}$$

where

$$\Theta = \begin{pmatrix} \exp(i\alpha/2) & 0 \\ 0 & \exp(-i\alpha/2) \end{pmatrix}_{\{l,r\}} \tag{39}$$

is the transformation matrix that accounts for the angular mismatch between the two mirrors. The transformation expressed in eq. (38), without term with $\Theta^{-1}$, arises because the reflected beam is expressed in a different basis from the incoming beam, since the *z* axis always points in the direction of propagation of light, which comes along with a reversal of the horizontal axis upon reflection to keep the coordinate axes right-handed [41]. The round-trip Jones matrix is given by [14]:

$$J_{TM} = \begin{pmatrix} r_1 r_2 \exp(2i\alpha) + \Delta r_1 \Delta r_2 & r_1 \Delta r_2 \exp(i\alpha) + \Delta r_1 r_2 \exp(-i\alpha) \\ r_1 \Delta r_2 \exp(-i\alpha) + \Delta r_1 r_2 \exp(i\alpha) & r_1 r_2 \exp(-2i\alpha) + \Delta r_1 \Delta r_2 \end{pmatrix}_{\{l,r\}},$$
(40)

which has *PT*-symmetric form, eq. (16), and the control parameter is :

$$\chi = \frac{r_1^2 r_2^2 \sin^2(2\alpha)}{r_1^2 \Delta r_2^2 + r_2^2 \Delta r_1^2 + 2 r_1 r_2 \Delta r_1 \Delta r_2 \cos(2\alpha)}. \tag{41}$$

Hence, the angle $\alpha$ between the two mirrors' principal axes can be used as the experimental control parameter to scan the exact symmetry region ($\chi \leq 1$) at small $\alpha$ values and the broken symmetry region at $\alpha > \alpha_{EP}$ (i.e., $\chi > 1$).

Important characteristics of the polarizations eigenstates of the Faraday and twisted-mode *PT*-symmetric resonators are shown in Fig. 3 as a function of their respective control parameter, $\varphi$ or $\alpha$. The eigenvalue spectra, in magnitude and phase are shown in Fig. 3(a) and 3(b): the two modes have a pure real eigenvalue spectrum in the neighborhood of $\varphi=0$ or $\alpha=0$, and the eigenvalues become a complex conjugate pair beyond a critical value $\varphi=\varphi_{EP}$ or $\alpha=\alpha_{EP}$. In the exact *PT* symmetry region, the magnitude of the two eigenvalues is different; hence, the polarization eigenstate with the larger magnitude will preferentially oscillate, while the other one will likely be suppressed due to gain saturation in the active medium, although SHB phenomena may allow the oscillation of the polarization mode with higher losses at pump power values well above threshold [15]. In the broken PT symmetry region, the magnitude of the two eigenvalues is the same, so dual polarization emission is expected, which can take the form of alternate oscillation of each mode (mode hopping) or simultaneous oscillation. The different phase, $\Delta\varphi$, of the two modes in the broken symmetry region implies a different oscillation frequency of the two modes.

The polarization eigenstate can be mapped to ($x,y,z$) coordinates on the Poincaré sphere, Fig. 3c and 3d. For each resonator, the state of polarization of the two modes is rectilinear ($z=0$) and rotates from $\{h,v\}$ states at $\varphi=0$ or $\alpha=0$ to a unique diagonal state ($x, y, z$) =(0, ±1, 0) at the EP, $\varphi=\varphi_{EP}$ or $\alpha=\alpha_{EP}$. At higher $\varphi$ or $\alpha$ values, each eigenstate becomes elliptical, with one axis of the ellipse oriented in the diagonal direction, i.e., $x=0$ and tends towards a circularly polarized state, $z=\pm1$, as the $\varphi$ (or $\alpha$) value is increased.

It is in the contrast of the standing wave that these two resonators differ the most, Fig. 3e and f. For the Faraday resonator, the polarization states of the two counter-propagating waves of each mode remain similar for any value of the $\varphi$ parameter. Hence, the intensity contrast of the standing wave remains close to the maximum value of one. On the other hand, for the twisted-mode resonator, the intensity contrast of the counterpropagating waves, which is maximum at $\alpha=0$, steadily decreases as $\alpha$ increases in the exact *PT* symmetry region; it reaches zero at the EP and remains at zero everywhere inside the broken *PT* symmetry region. It is thus possible to eliminate SHB by operating a twisted-mode *PT*-symmetric laser in that region. Hence, operating such a laser near an EP is a privileged point of operation for achieving true single mode emission from a microchip laser, because both dual polarization emission and spatial hole burning will likely be suppressed [14,15].

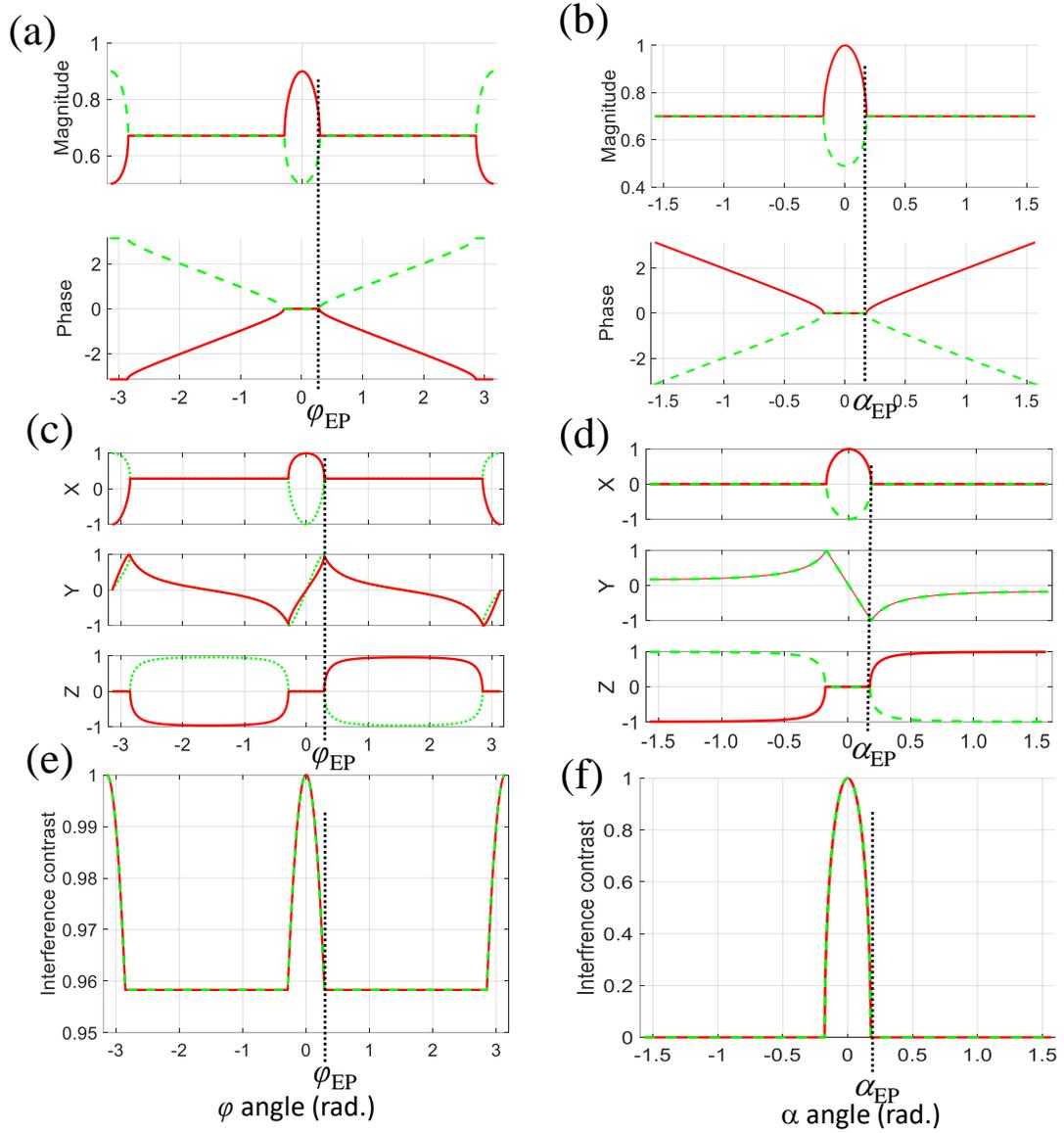

Fig. 3. Important characteristics of the eigenstates of the Faraday (a), (c), (e) and twisted-mode (b), (d), (f) *PT*-symmetric resonators as a function of $\varphi$ or $\alpha$. For each mode, (a) (b): the magnitude and phase of the eigenvalues, (c) (d): the coordinates localization of the polarization eigenstates on the Poincaré sphere, (e) (f): the contrast of the standing wave are shown. For the Faraday resonator, the parameters are: $\Delta r = 0.2$ and $r = 0.7$. For the twisted-mode resonator, they are $\Delta r_1 = \Delta r_2 = 0.15$ and $r_1 = r_2 = 0.85$, with identical mirrors. The two eigenmodes are ploted with red solid and green dashed lines.

# 5. DISCUSSION AND CONCLUSION

5.1 Connection between *PT*-symmetric with antiunitary-symmetric operators.

The *P* and *T* operators were defined in the most intuitive way in Fig. 1: *P* is the inversion of space and *T* is the reversal of motion. With these definitions, $J_{PT}$ has the specific form of eq. (16) in the $\{l,r\}$ basis with four free parameters, but can take the general form shown in eq. (17) with six free parameters by choosing a different orthogonal basis, $\{\vec{u}_1, \vec{u}_2\}$, parametrized as:

$$\vec{u}_1 = \left(\cos\theta/2 \quad , \quad \sin\theta/2 \exp(i\varphi)\right)^T_{\{l,r\}} \tag{42a}$$

and

$$\vec{u}_2 = \exp(i\varepsilon)\left(-\sin\theta/2 \quad , \quad \cos\theta/2 \exp(i\varphi)\right)^T_{\{l,r\}}, \tag{42b}$$

where $\theta$ and $\varphi$ are the polar angle and the longitude on the Poincaré sphere, and $\varepsilon$ is an arbitrary phase factor. Whatever the chosen basis, the eigenvalues are the same. It is thus possible to generalize the *PT*-symmetric operator in polarization space by redefining the *P* and *T* operators in such a way that their representation that we found in one specific basis does in fact correspond to another basis.

For example, let us represent new operators *P'* and *T'* in the $\{h,v\}$ basis respectively as:

$$P'_{\{h,v\}} = \begin{pmatrix} 0 & 1 \\ 1 & 0 \end{pmatrix} \tag{43}$$

and

$$T'_{\{h,v\}} = C, \tag{44}$$

which are the actual *P* et *T* representations found in the $\{l,r\}$ circular basis. We then find that the *P'* and *T'* operators behave quite differently from *P* and *T*, as shown in Fig. 4: *P'* does reverse the handedness of the polarization as expected for a parity operator, but it also mirrors the ellipse of polarization around the plane bisecting the *x,y* axes. The *T'* operator reverses the rotation direction without reversing the ***k*** vector; this is perplexing because this causes a reversal of the handedness of the polarization, in stark contrast with the true time reversal operator *T*, which does not. Yet, although the *P'* and *T'* operators are different from *P* and *T*, operators $J_{P'T'}$ commuting with *P'T'* will share the same spectrum as $J_{PT}$ because $J_{P'T'}$ and $J_{PT}$ differ only by a unitary transformation. Hence, the notions of broken and exact symmetries apply to these operators just as well. In this example, the structure of $J_{P'T'}$ will be:

$$J_{P'T'\{h,v\}} = \begin{pmatrix} \beta+i\gamma & \delta+i\eta \\ \delta-i\eta & \beta-i\gamma \end{pmatrix}, \tag{45}$$

in the {$h,v$} basis instead of the {$l,r$} basis, as in eq. (16).

Therefore, the concept of *PT* operator in polarization space can be extended to any pair of operators that are equivalent to *PT* by a unitary transformation. Such *P'T'* operator is just an arbitrary antiunitary Jones Matrix, whose general form is the association of a unitary matrix with the complex conjugate operator, as in eq. (8). This fact that an operator commuting with an antiunitary operator must either have a real spectrum or the latter appears in complex conjugate pairs was pointed out by Bender et al. [42]. Hence, the concepts of broken and exact symmetries also exist for Jones matrices that commute with an antiunitary operator. Such Jones matrices still take the form of eq. (17). We thus conjecture that there exists a larger class of resonator geometries that can probe the exact and broken symmetries, as well as the EP at the transition between the two regions, that are ''antiunitary-symmetric'' rather than PT-symmetric and that are yet to be discovered.

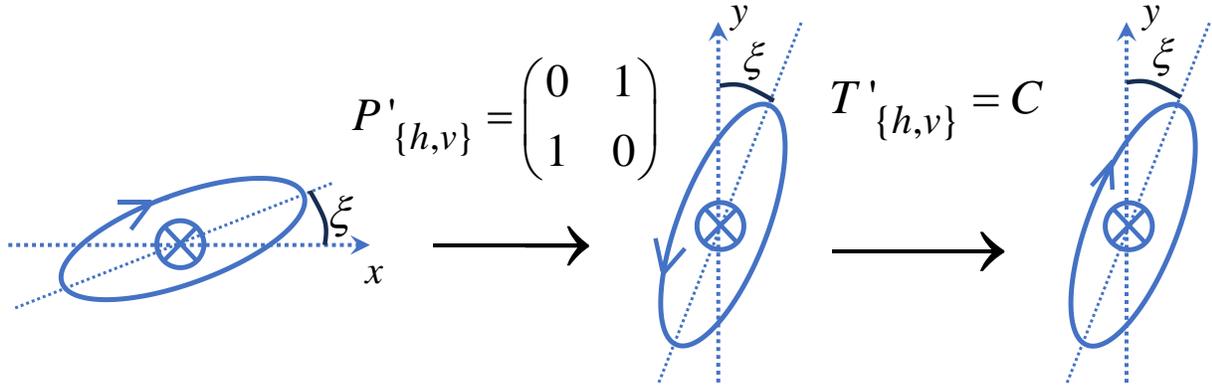

Fig. 4. Behavior of *P'* and *T'* operators, defined on the figure, on a given state of polarization. The *T'* and *PT* operators bear little resemblance with the actual *P* and *T* operators but the P'T'-symmetric operator, $J_{P'T'}$, has the same spectrum as $J_{PT}$.

5.2 $J_{PT}$ englobes all Hermitian and unimodular unitary matrices.

It is interesting to consider the *PT*-symmetric Jones matrices in the broader context of polarization optics. We generally divide polarization optical components into two families: the polarizers, which have different values of transmittance for orthogonal states, and the retarders, which have different phase delays between orthogonal states.

Pure polarizers are described, within a global phase factor, by Hermitian matrices, *H*; they show different values of the modulus of the transmission (or reflection) coefficient for orthogonal Jones states, without relative phase shifts between them. In any orthogonal basis of Jones vectors, they have the property of being self-adjoint:

$$H^\dagger = H. \tag{46}$$

Any Hermitian Jones matrix is also *PT*-symmetric. This can be seen by taking $B=0$ and $\varepsilon=0$ in the general form, eq. (17):

$$H = J_{PT}\big|_{B=0,\varepsilon=0} = \begin{pmatrix} A+C\sin\theta & C\cos\theta + iD \\ C\cos\theta - iD & A-C\sin\theta \end{pmatrix} \equiv \begin{pmatrix} E & G+iD \\ G-iD & F \end{pmatrix}, \qquad (47)$$

where arbitrary real values of *E*, *F* and *G* can be obtained from suitably chosen real values of *A*, *C*, and $\theta$. We just saw in section 5.1 that there exists an orthogonal basis, $\{\vec{u}_1, \vec{u}_2\}$, where $J_{PT}$ takes the form:

$$J_{P'T'\{\vec{u}_1,\vec{u}_2\}} = \begin{pmatrix} \beta + i\gamma & \delta + i\eta \\ \delta - i\eta & \beta - i\gamma \end{pmatrix}. \qquad (48)$$

This basis is $\{l,r\}$ when *P* and *T* are defined in the conventional way. Since this matrix is Hermitian for $\gamma=0$, we see from eq. (25) that the $\chi$ value corresponding to the Hermitian matrix is $\chi=0$, which could be described as *extreme* exact *PT* symmetry. Note that the identity matrix, as well as a real multiple of it, is Hermitian with undefined $\chi$ values.

The second category, the retarders, corresponds, to unitary elements, *U*, wherein:

$$U^{-1} = U^\dagger. \qquad (49)$$

There exists an orthogonal basis that diagonalizes *U* as:

$$U = \exp(i\alpha)\begin{pmatrix} \beta + i\gamma & 0 \\ 0 & \beta - i\gamma \end{pmatrix}, \qquad (50)$$

where $\beta^2 + \gamma^2 = 1$. If we set the global phase shift $\alpha$ to 0 or $\pi$, then *U* has a determinant equal to one and also has *PT*-symmetric form with $\delta^2 + \eta^2 = 0$, which implies that $\chi \to \infty$, which is an *extreme* case of broken PT symmetry.

In summary, polarizers and retarders are PT-symmetric within a global phase factor; Hermitian ($\chi=0$) and unitary matrices with unit determinant ($\chi \to \infty$) are at opposite ends of the spectrum of the broken and exact symmetries. These results are summarized in a Venn diagram in Fig. 5. As explained in section 5.1, this diagram applies equally well for Jones matrices that commute with antiunitary operators, because the latter are unitarily similar to *PT*-symmetric Jones matrices.

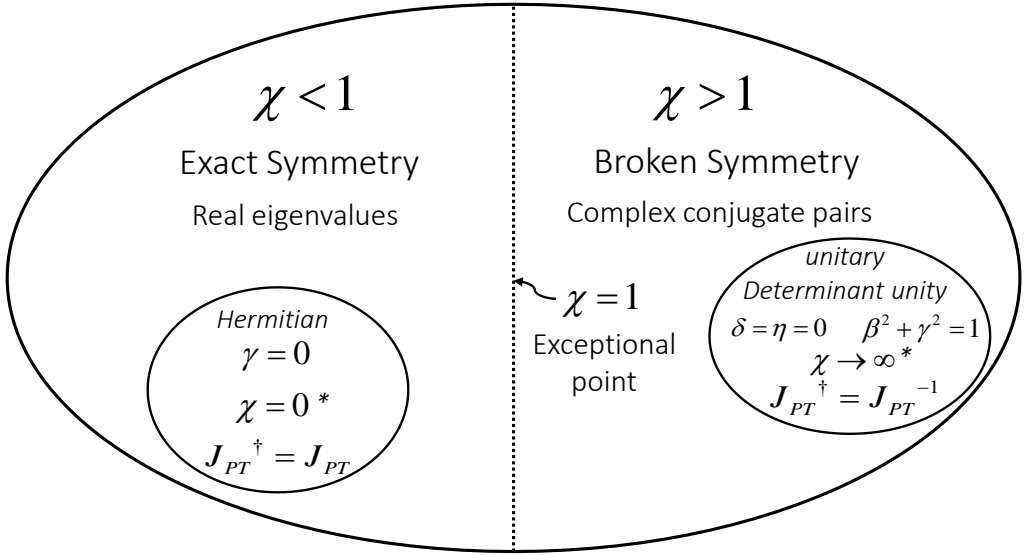

*: except the identity or proportional to identity matrix: indefinite χ.

Fig. 5. Venn diagram of PT-symmetric matrices. Hermitian matrices and unimodular unitary matrices are special cases of exact and broken symmetry, respectively.

5.3 Concluding remarks

In summary, we defined a parity reflection - time reversal (*PT*) operator for arbitrary polarized pure states of light. We then established the form of Jones matrices, $J_{PT}$, that commute with *PT*. We showed that $J_{PT}$ can display either real or complex conjugate eigenvalue spectra depending on the value of a control parameter χ, and the two eigenstates merge to a single state at the transition between the two regions. We traced the origin of this peculiar phenomenon to the fact that the *PT* operator is antilinear. We described *PT*-symmetric laser resonators, which are concrete experimental examples of *PT*-symmetric Jones matrices that possess a control parameter that enable one to experimentally cover the exact and the broken *PT*-symmetric regions. One resonator, the twisted mode resonator shows a null contrast of the standing wave in the broken *PT* symmetry region, which eliminates spatial hole burning in microchip lasers and makes it suitable for single mode emission. Finally, by using unitary transformations, we extended the concept of *PT* operator to other similar operators that differ appreciably from the PT operator but retain all the features of broken and exact PT-symmetry. We showed that Hermitian and unitary operators with determinant equal to one are special cases of PT-symmetric Jones matrices, the former belonging to the exact, the latter to the broken PT-symmetry region.

**APPENDIX A: PT-SYMMETRY IN JONES MATRICES VERSUS QUANTUM SYSTEMS.**

We point out one important difference between the eigenvalue spectrum of *PT*-symmetric Jones matrices and that of a *PT*-symmetric quantum system. In quantum mechanics, a two-level quantum system is generally described by the Schrödinger equation which, in the matrix formalism, is given by the following equation:

$$i\hbar \frac{d}{dt}\begin{pmatrix} u_1 \\ u_2 \end{pmatrix} = \begin{pmatrix} H_{11} & H_{12} \\ H_{21} & H_{22} \end{pmatrix}\begin{pmatrix} u_1 \\ u_2 \end{pmatrix}, \tag{A1}$$

where $u_1$ and $u_2$ are two linearly independent states. Such equation is used to represent, e.g., coupled systems, such as waveguides displaying both gain and loss. The diagonal terms may represent the oscillation frequency in the absence of coupling and non-diagonal elements are coupling coefficients. The diagonalization of *H* yields eigenstates and eigenvalues, the real part of which is the eigenenergy or emission frequency and the imaginary part describes loss or gain.

It is possible to express the evolution of the polarization of a laser resonator with an equation similar to (A1). From eq. (27), we have

$$\vec{u}^{[n+1]} - \vec{u}^{[n]} = (J_{RT} - I)\vec{u}^{[n]}. \tag{A2}$$

If we perform a coarse-graining approximation, where the intracavity round-trip time, $T_{RT}$, is assumed to be much smaller than other relevant time scales, the evolution of $\vec{u}$ as a function of time at a fixed position inside the resonator will approximately follow:

$$\frac{d\vec{u}}{dt} \simeq \frac{(J_{RT} - I)}{T_{RT}}\vec{u} \triangleq M\vec{u}, \tag{A3}$$

which is similar to (A1). Note that the eigenstates of $J_{RT}$ and *M* in (A3) are the same. One could analyze the eigenvalues, $\lambda_{Mi}$, of *M* instead of those of $J_{RT}$ and would find that the real part of $\lambda_{Mi}$ describes loss or gain, while the imaginary part describes the phase shift after a roundtrip. Although this coarse graining approximation was not needed for our analysis, such view is helpful to highlight a key difference between (A1) and (A3), namely that the imaginary *i* is absent in the left-hand side for the latter. As a result, the interpretation of eigenmode emission frequency and loss in the exact and symmetry is opposite for $J_{RT}$ compared to *H*, as shown in Table AI. However, the general form of a two-by-two PT-symmetric Hamiltonian remains that outlined for the Jones matrix in (17).

Table AI. Differences in the interpretation of the eigenvalue spectrum of two-dimensional *PT*-symmetric Hamiltonian and Jones matrices of a laser resonator.

| *PT* symmetry | Eigenvalues | Two-level quantum system (Eq. (A1) | Jones matrix of a laser resonator |
|---|---|---|---|
| Exact | Pure real | Distinct emission frequencies Same loss/gain | Same emission frequency Different loss/gain |
| Broken | Complex conjugate pair | Same emission frequency Different gain/loss | Distinct emission frequencies Same loss/gain |

## ACKNOWLEDGMENT

The author acknowledges financial support from the Natural Sciences and Engineering Research Council of Canada Discovery grant program, the New Brunswick Innovation Foundation and the Canada Foundation for Innovation.